\documentstyle[prl,aps,epsfig]{revtex}
\begin{document}

\title{Estimating Stochastic Gravitational Wave Backgrounds
with  Sagnac Calibration}
\author{Craig J. Hogan}
\address{Astronomy and Physics  Departments, 
University of Washington,
Seattle, Washington 98195-1580}
\author{Peter L. Bender}
\address{JILA, University of Colorado and National Institute of Standards
                and Technology, Boulder, Colorado 80309-0440}
\maketitle
\begin{abstract}
Armstrong {\it et al.} have recently  presented   new ways of
combining signals to  precisely cancel laser frequency noise in spaceborne interferometric
gravitational wave detectors such as LISA\cite{armstrong,tintoarmstrong,estabrook,tinto}.   
One of these combinations, which we will call the ``symmetrized Sagnac observable'',    is
much less sensitive to external signals at low frequencies than other combinations, and thus
can be used to determine the instrumental noise level\cite{tinto}. 
We note here that this  calibration of the instrumental noise permits smoothed versions of the
power spectral density of stochastic gravitational wave backgrounds to be determined with
considerably higher accuracy than earlier estimates, at frequencies where
one type of noise strongly dominates and is not substantially correlated
between the six main signals generated by the antenna. We illustrate this technique by
analyzing simple estimators of  gravitational wave background power, and show 
that     the instrumental sensitivity to broad-band backgrounds   at some frequencies can be
improved by a significant  factor of as much as $  ( f\tau/2)^{1/2}$ in spectral density
$h_{rms}^2$ over the standard method, where $f$ denotes  frequency and  $\tau$ denotes
integration time, comparable to  that which would be  achieved by cross-correlating two
separate antennas.
 The applications of this approach to studies of
astrophysical gravitational wave backgrounds generated after recombination and to
searches for a possible primordial background are discussed. 
 With appropriate mission design, this
technique   allows an  estimate of the cosmological background from
extragalactic white dwarf binaries and
  will enable LISA to reach  the astrophysical
 confusion
noise of  compact binaries from about 0.1 mHz to about $ 20 $mHz. 
In a smaller-baseline
follow-on mission, the technique allows several
orders of magnitude improvement in  sensitivity 
to primordial backgrounds up to about 1 Hz.
\end{abstract}

\section{Introduction}

In ground-based interferometric gravitational wave detectors such as LIGO,
VIRGO, GEO-600 and TAMA, stochastic backgrounds will be best detected by
correlating signals from more than one interferometer within a wavelength of
each other.  If in-common noise sources can be eliminated, the correlation
allows a direct estimate of the ``noise" coming from gravitational waves,
separately from instrumental sources of noise. In this way
the detection
of a broadband background can take
advantage of a broad detection bandwidth $B$, and sensitivity
to rms strain in a broad band  
grows with time like $h_{rms}\propto (B\tau)^{-1/4}$.

The problem is different for the Laser Interferometer Space Antenna (LISA)\cite{lisaref},
which consists of three spacecraft in a triangle configuration. Although two ``independent''
observables can be measured
with this arrangement, yielding orthogonal polarization information for sources,
the  observable signals are not  truly independent since they   include correlated
instrumental
noise.  Separation of  the instrumental
noise from  stochastic gravitational wave signals requires 
an alternative approach, as well as careful attention
to correlations in the different types of noise affecting the signals. 

A fundamental recent development has been the introduction by Armstrong {\it et
al.}
of a new way of precisely cancelling laser frequency noise in
interferometric
gravitational wave detectors where the arm lengths are not exactly equal
\cite{armstrong,tintoarmstrong,estabrook}.  It is shown in these papers that
for a triangular geometry
 as is used in LISA, the
signals from detectors in the different satellites 
can be combined, if the hardware allows, to give various observables that
are free of the
laser frequency noise. 
In addition, several of these observables have considerably reduced
sensitivities to gravitational wave signals at low frequencies (below about
30 mHz, corresponding to the 33-second roundtrip light travel time on one
arm
of the triangle).  The observables $\alpha, \beta$ and $\gamma$
defined
in \cite{armstrong} correspond to  Sagnac observables:
they correspond  to taking the difference in phase for laser beams that have
gone around the triangle in opposite directions, each starting from a different
spacecraft.

  However, another of the observables defined in \cite{armstrong}, called $\zeta$, has even
less sensitivity to gravitational waves at low frequencies.  We will refer
to
it as the ``symmetrized Sagnac observable", since the signals that are
combined
to form $\zeta$ are the same as for $\alpha, \beta$ and $\gamma$, but they are
evaluated
at very nearly the same time instead of at substantially different times.
This observable allows a more complete ``switching off'' of the sky signal, and can
be
used to give a valuable determination of the other sources of noise in the
interferometer, as discussed in \cite{armstrong,tintoarmstrong,estabrook}.

More recently, Tinto et al.\cite{tinto} have discussed the problem of separating the
confusion noise due to many unresolvable galactic and extragalactic binaries
in each frequency bin from instrumental noise.  In particular, they consider
the case where the confusion noise level is comparable with or larger than
the
instrumental noise level.  They  show  that what we are calling the
symmetrized Sagnac observable permits
the confusion noise level to be established reliably.

What apparently has not been pointed out previously is that using the symmetrized Sagnac
observable to calibrate the instrumental noise potentially makes possible
considerably  
higher sensitivity for determining smoothed values of the power spectral
density
for broad-band isotropic gravitational wave backgrounds, such as 
binary confusion backgrounds or primordial stochastic backgrounds.  
For each frequency bin, with a width roughly equal to the inverse of the
data
record length, the noise power in the  sky signal can be separated from the
instrumental noise power by
using estimators which combine the symmetrized Sagnac
observable with the other observables (such as Michelson observables)
which are fully sensitive to gravitational waves.  For isotropic
backgrounds with  fairly smooth spectra, the precision can be improved
substantially
by    integrating the estimated power spectral density   over many 
spectral bins, after removing and fitting out recognizable 
binary sources.
 In this paper we discuss  this idea and its impact on studies of
 backgrounds observable by LISA and  by a possible high-frequency follow-on mission.
Our main conclusion is that this capability should be included in 
science optimization studies for the detailed mission design for LISA.

\section{ESTIMATING STOCHASTIC BACKGROUNDS USING THE SYMMETRIZED SAGNAC OBSERVABLE  AND A
BROAD BANDWIDTH}

For simplicity, we will consider only the 6 Doppler signals $y_{ij}$
($i,j=1,2,3$) introduced by Armstrong
et al. in \cite{armstrong}, rather than the  more complete results 
including the additional 6 signals $z_{ij}$  given by Estabrook et al.
\cite{estabrook} to allow for having two separate
proof masses in each spacecraft. 
This corresponds to setting the $z_{ij}$ in \cite{estabrook}
equal to zero.  The two lasers in each
spacecraft are thus assumed to be perfectly phase locked together, but to run
independently of the lasers in the other two spacecraft.  On each spacecraft, the
phases of the beat signals between the laser beams from the two distant spacecraft
and the local lasers are measured
as a function of time and recorded.  This gives the total of 6 signals that
are considered.  They are sent to a common spacecraft and then combined,
with
time delays equal to the travel times over different sides of the triangle,
to
give various different observables that are free of the phase noise in the
lasers. 

The data combinations relevant for this discussion are illustrated
in Figure 1. 
 Although laser and optical bench noise
exactly cancel in these combinations, they contain various  mixtures of gravitational
wave signals and instrumental noise.    
The   Sagnac observables
 $\alpha$, $\beta$ and $\gamma$ have a lower sensitivity for
gravitational waves at   frequencies near 1 mHz than do the 
Michelson observables $X$, $Y$, $Z$   discussed
by Armstrong et al. Thus, we will base our strategy on using
 the observables $X, Y$ and
$Z$ to detect the gravitational waves, and the 
symmetrized Sagnac observable $\zeta$ to calibrate the noise.

 The technique usually considered for estimating the stochastic background is to use time
variations in the observed power during the year to model out noise sources with
non-isotropic components such as confusion noise from galactic binaries. (Integration over a
year   will give a nearly isotropic response to incoming gravitational wave
power\cite{giampieri,giampieripolnarev,giampieripolnarev2}.)
 However, the uncertainty in the
instrumental noise level still remains.  The sensitivity is then limited to 
a factor of order unity times that
obtainable in one frequency resolution element, $ \delta f \approx \tau^{-1}$,
where $\tau$ is the length of the time series.
This factor is the fractional
uncertainty in the level of the instrumental noise. In effect this means that
the sensitivity to stochastic backgrounds does not increase with time.
The technique we propose is to use $\zeta$ to
calibrate
the noise power levels differentially in each frequency bin,
i. e. use Sagnac calibration, allowing a sum of 
sky signal from  a broad bandwidth, $B\approx f/2$. For
broadband backgrounds, 
this approach begins to ``win''
after an integration time $\tau\approx 2 f^{-1}$.

We now sketch in more detail a specific strategy  for analysing the data.  This strategy
allows an accurate calibration of the main sources of noise entering into the $y_{ij}$,
without assuming that these either are the same or are known a priori.  We adopt the
notation of \cite{estabrook}, and assume that the complex Fourier coefficients $X_k$, $Y_k$,
$Z_k$ and
$\zeta_k$ for $X$, $Y$, $Z$ and $\zeta$ have been derived from a long data set, such as
perhaps a year of observations.  We also define $\eta_k^2$ as  the mean of the squares of the
absolute values of
$X_k$, $Y_k$, and $Z_k$.

Since the laser noise $C_{ij}$  exactly cancels for 
these combinations, the main instrumental noise sources are due to the noninertial
changes in the velocities  of the proof masses $\vec v_{ij}$ (the ``proof mass noise''
$y_{ij}^{proof\ mass}$)
and the combination of noise from pointing errors, shot noise,
and other optical path effects (the ``optical path noise''
$y_{ij}^{optical\ path}$).
 From eqs. 3.5
and 3.6 in \cite{estabrook}, and from cyclic perturbations of eqs. 3.1 and 3.2, with
the $z_{ij}  = 0$, the noise power spectral densities $S_X(f)$, $S_Y(f)$
$S_Z(f)$ and $S_\zeta(f)$, without gravitational waves, can be obtained
in terms of the  $y_{ij}^{proof\ mass}$ and $y_{ij}^{optical\ path}$. 
We define $S_{ave}(f)$ to be the average of $S_X(f)$,
$S_Y(f)$ and $S_Z(f)$,
 and $\langle S_y^{proof\ mass}\rangle$ and
 $\langle S_y^{optical\  path}\rangle$   to be the averages
 of the corresponding noise power spectral
densities for the 6 signals $y_{ij}$.
Then we obtain the same results as for eqs. 4.1 and 4.3 in 
\cite{estabrook}, except with
$S_X(f)$ replaced by $S_{ave}(f)$
 and with $S_y^{ proof\ mass}$ and
$S_y^{ optical\ path}$ replaced with their average values:
\begin{equation}\label{eqn: alphanoise}
S_{ave}(f)=
[ 16\sin^2(2\pi f L)]\left\{ [2\cos^2(2\pi f L)+2]
\langle S_y^{proof\ mass}\rangle+
 \langle S_y^{optical\  path}\rangle\right\}
\end{equation}
\begin{equation}\label{eqn: zetanoise}
 S_\zeta(f)=
24\sin^2(\pi f L)
\langle S_y^{proof\ mass}\rangle
+6\langle S_y^{optical\  path}\rangle.
\end{equation}
These formulas do not assume that the
noise contributions to the  individual $y_{ij}$ are the same.

The quantities $X_k$, $Y_k$,
etc., can be divided into an instrumental noise part and a gravitational
wave background part:  i.e. $X_k = X_{n,k} + X_{GW,k}$, etc.  We
can then define  an estimator  $E_k$ for the gravitational
wave  power,
\begin{equation}
        E_k = \eta_k^2
                - D(f)|\zeta_k|^2, 
\end{equation}
where
\begin{equation}
        \eta_k^2 \equiv (1/3)[|X_k|^2 + |Y_k|^2 + |Z_k|^2].
\end{equation}
The coefficient $D(f)$ is defined in such a way that 
the noise component of the second term subtracts  (on average)  the  noise
component of the first, leaving only contributions
from the gravitational wave power of both terms; that is, 
  $\langle E_k\rangle $ is a  known multiple of the GWB.
In general $D(f)$ will be computed numerically based on a  model of LISA
and its noise sources.
Here we  estimate the bias  in the estimate
and the sensitivity level for a detection or upper limit based on $E_k$,
for two situations where we can identify analytical approximations to 
$D(f)$ based on the simple model described above.

We first  define a high-frequency 
estimator $E_k$ for the GW background power, useful when the 
optical path noise dominates (but when $f$ is not so high that the Sagnac
combination becomes nearly fully sensitive to gravitational waves):
\begin{equation} \label{eqn: estimator}
        E_k = \eta_k^2
                - [S_{ave}/S_\zeta]_{est} |\zeta_k|^2,
\end{equation}
For frequencies high enough so that
\begin{equation}
        R = \langle S_y^{proof mass}\rangle /\langle S_y^{optical path}\rangle 
\end{equation}
is small,
\begin{equation}
        [S_{ave}/S_\zeta]_{est} = G_1(f)[(8/3)\sin^2(2 \pi fL)],
\end{equation}
where
\begin{equation}
        G_1(f) =   {1+2[1 + \cos^2 (2 \pi f L)]R_{est}\over  1+4[\sin^2 (\pi f L)]R_{est}}.
\end{equation}
To first order in the actual value of $R$ minus the estimated value $R_{est}$,
the bias in $E_k$ is given by
\begin{equation}
        (\delta E_k)_{bias} =
            [S_{ave}(f)][2 + 2\cos^2(2 \pi fL) - 4 \sin^2(\pi fL)][R - R_{est}].
\end{equation}

   To the extent that the bias in $E_k$ can be neglected,
$\langle E_k\rangle $ depends just on the GWB power:
\begin{equation}
        \langle E_k\rangle  = S_{GW,ave} - [(8/3)\sin^2(2 \pi fL)]S_{GW,\zeta}
\end{equation}
where $S_{GW,ave}$ is defined as
\begin{equation}
        S_{GW,ave} = \langle \eta_{GW,k}^2\rangle 
\end{equation}
and 
\begin{equation}
        S_{GW,\zeta} = \langle |\zeta_{GW,k}|^2\rangle  = \epsilon S_{GW,ave}.
\end{equation}
At frequencies which are not too high, the Sagnac gravitational wave sensitivity 
is low so $\epsilon<<1$.
The estimator is most useful at frequencies low enough so that $\langle E_k\rangle $ is
comparable with $S_{GW,ave}$ and thus can be used to estimate the GW
background power efficiently.  For LISA with triangle sides of length
$L = 5\times 10^6$ km, or 16.67 seconds in units with $c = 1$, this condition is
satisfied if $f \le f_{crit} \approx 25$ mHz.

   The sensitivity to GWB is given by estimating the uncertainty
$\delta E_k$ in $\langle E_k\rangle $ from the relation
\begin{equation}
         \delta E^2 = \langle E_k^2\rangle  - \langle E_k\rangle^2.
\end{equation}
This is done in the appendix.  The results are found to depend on the
individual noise spectral densities for the six main LISA signals, rather
than just their average value.  For the case of all six noise spectral
densities being equal, the results are:
\begin{equation}
        \delta E^2 \approx [(64/3)\sin^4(\omega L)]
           [9 + 4\cos(\omega L) - \cos(2 \omega L)]\langle S_y^{optical path}\rangle^2,
\end{equation}
\begin{equation}
        \delta E\leq [16 \sin^2(\omega L)]S_y^{optical path}.
\end{equation}
If we let $\delta E =  \lambda \langle \eta_{n,k}^2\rangle$, then
the factor $\lambda $ characterizes the noise level of the estimate
relative to an ideal instrument-noise-limited measurement.

   Similarly, for very low  frequencies where $R >> 1$ and the proof mass
noise dominates, the estimator becomes
\begin{equation}
        E_k = \eta_k^2
                -G_2(f)[(16/3)\cos^2(\pi fL)][1 + \cos^2(2 \pi fL)]|\zeta|^2
\end{equation}
where
\begin{equation} \label{eqn: G2}
        G_2(f) =  {1+R^{-1}_{est}[2+2\cos^2 (2 \pi f L)]^{-1}\over
                         1+R^{-1}_{est} [4\sin^2 (\pi f L)]^{-1}}.
\end{equation}
In this case, for all six noise spectral densities equal, $\delta E_k$ can
be shown to be
\begin{equation}
        \delta E_k \approx [(2/3)+(4/3)\cos\ (\omega L)+(5/6)\cos^2(\omega L)]^{1/2}
                        \langle S_y^{proof mass}\rangle.
\end{equation}
The interesting frequency range with $R >> 1$ is near $100$ microhertz, and
thus $\omega L$ is very small, giving
\begin{equation}
        \lambda \approx (17/6)^{1/2} = 1.7.
\end{equation}
It should be noted that, for frequencies $f \stackrel{<}{\sim} 100$ microhertz, the
combination $[(R_{est})^{- 1}]/4 \sin^2(\omega L/2)$ in the denominator of 
Equation (\ref{eqn: G2})
is expected to be small for LISA even though $(\omega L)$ is very small.  Thus $G_2(f)$ will
be very close to unity, and the bias in $E_{k,n}$ is negligible.

   The standard estimate of the amplitude signal-to-noise ratio $S/N$ for
detecting a gravitational wave background is given by
\begin{equation}
        (S/N)_k^2 = {S_{GW,X}\over\langle |X_{n,k}|^2\rangle}.
\end{equation}
As noted above, this sensitivity estimate implicitly assumes that the uncertainty in
estimating the instrumental noise power level is about the same as the
level itself.  However, the error in estimating the instrumental noise
level may well be highly correlated over a bandwidth comparable with the
frequency, so that averaging the results from many frequency bins gains
little if anything.

   With the symmetrized Sagnac calibration approach, the $S/N$ contributions
from individual frequency bins are given by
\begin{equation}
        (S/N)_k^2 = \{1 - [(8/3)\sin^2(\omega L)]\epsilon\}
                        { S_{GW,ave} \over  \lambda\langle \eta_{n,k}^2\rangle}.
\end{equation}
Thus there are two possible inefficiency factors, characerized by $\epsilon$ and
$\lambda$. 
However, these are more than offset  for detecting broad-band backgrounds,
since   
the   contributions from individual frequency bins can now be averaged over
a bandwidth of roughly $f/2$ to give an
 improvement in $(S/N)^2$ by a factor of
about
$(f \tau/2)^{1/2}$.  Since $S_{GW,ave} = S_{GW,X}$ for an isotropic background,
the overall reduction in the rms background gravitational wave amplitude
needed in order to achieve $S/N = 1$ can be as large as a factor
\begin{equation}
        F = \{1 - [(8/3)\sin^2(\omega L)]\epsilon\}^{1/2}
                \left[{f \tau\over 2 \lambda^2}\right]^{1/4},
\end{equation}
relative to the standard estimated sensitivity.
 The symmetrized Sagnac calibration approach achieves
about the same gain in sensitivity as the cross-correlation approach
employed by ground-based experiments, and discussed by Ungarelli and
Vecchio \cite{ungarelli} for two separate LISA-type space-based antennas.

The discussion above has implicitly
assumed that the dominant instrumental noise contributions to all of the six recorded
signals $y_{ij}$ are  not correlated in phase.  This is certainly true for the
shot
noise, but careful instrumental design will be necessary to make it a useful
approximation for other noise sources.  For example, wobble of the pointing
of
a given spacecraft could give rise to correlated noise in the received
signals
at the other two spacecraft due to wavefront distortion.  Also, correlated
proof mass acceleration noise for two proof masses on the same spacecraft
can
occur if the effect of common temperature variations is significant.  A
quantitative discussion of such correlations will be required before the
extent of realistically feasible improvements in stochastic background
measurements can be determined.  However, this is beyond the scope of our
current knowledge of such effects.  We therefore will assume that the six
signals may have different noise levels but are uncorrelated in phase.  Our
results thus are rough upper limits to the possible improvements with the
symmetrized Sagnac   calibration method.

\section{ SENSITIVITY LIMITS AND BINARY BACKGROUNDS }

The approximate threshold sensitivity of the planned LISA antenna with $5\times 10^6$ km arm
lengths and for a signal-to-noise ratio S/N = 1 is shown in Figure 2.  The
sensitivity using 
the standard Michelson observable    can
be
approximated by a set of power law segments:
\begin{eqnarray}
h_{rms } & = & 1.0\times 10^{-20} [f/10 mHz]/\sqrt{\rm Hz},\qquad  10 mHz < f \\ \nonumber
& & 1.0\times 10^{-20}/\sqrt{\rm Hz},\qquad 2.8 mHz < f < 10 mHz\\ \nonumber
& & 7.8\times 10^{-18} [(0.1 mHz/f)^2]/\sqrt{\rm Hz}, \qquad 0.1 mHz < f < 2.8mHz\\ \nonumber
& & 7.8\times 10^{-18} [(0.1 mHz/f)^{2.5}]/\sqrt{\rm Hz}, \qquad 0.01 mHz < f<0.1mHz   
\end{eqnarray}
where   the sensitivity has been averaged over the
source
directions.  Below 100 $\mu$Hz there is no adopted mission sensitivity
requirement, but the listed sensitivity has been recommended as a goal for
frequencies down to at least 10 $\mu$Hz, provided that the cost impact is
not
too high.

A number of authors have discussed the expected levels of gravitational wave
signals due to binaries in our galaxy, and the essentially isotropic
integrated background from all other galaxies out to large red shifts.
The
normalization is uncertain, since only a few   binaries
 above 0.1 mHz in frequency are known, since they were selected
from highly biased surveys, and since the evolutionary history for some is
poorly constrained.   We adopt most of the
levels estimated in \cite{hils90} for the total binary
backgrounds,  with estimates from \cite{benderhils} for
  the reduction of confusion noise at higher frequencies
by fitting out Galactic binaries.  
The estimate for helium cataclysmics discussed in \cite{hilsbender}
is not included.

For close white dwarf
binaries (CWDBs), a factor  of 10 lower space density than the maximum yield
 estimated earlier from
models     of stellar  populations 
(e.g. \cite{webbink}) is assumed.
However, the resulting value is within a factor 2 of the
latest theoretical estimate of Webbink and Han \cite{webbink2}.
The factor 10 reduction factor is conventional, as discussed near the end of \cite{hils90}, 
and gives a 
signal level  a factor
$10 ^{1/2}$ lower than given in Table 7 of \cite{hils90}.  It should be noted that
there is  
about a factor of three uncertainty in the estimated total galactic signal
level,
and the estimated extragalactic signal level is even more uncertain.  The ratio of
extragalactic to galactic signal amplitudes is taken to be 0.2 for CWDBs and
0.3 for neutron star (NS) binaries.  (A ratio of 0.3 was found by Kosenko and
Postnov\cite{kosenko} for CWDBs with an assumed history of the star formation rate
and
for cosmological parameters $\Omega_{tot} = 1$ and $\Omega_\Lambda = 0.7$.  However,
the value of 5.5 kpc that they used for the scale height of the distribution
perpendicular to the plane of the disk is more appropriate for neutron
stars,
and a reduction by a factor of about 1.5 is needed for a CWDB scale height
near 90 pc.)

Below about 1 mHz there are so many galactic binaries that there will be many
 per frequency bin for one year of observations, and only a few of the
closest ones can be resolved.  Above roughly 3 mHz most Galactic binaries
will
be a few frequency bins apart, and can be solved for despite sidebands due to
the motion and orientation changes of the antenna.   
The
  effective spectral amplitude of
the confusion noise from both galactic and extragalactic
binaries remaining
after the resolved binaries have been fitted out of the data record 
(see e.g. \cite{benderhils})
is shown in Figure  2.  Essentially
none of the extragalactic stellar-mass
binaries can be resolved with LISA's sensitivity (in contrast to intense signals from 
an expected small number involving massive black holes (MBHs)).

Except for the shot noise, it is difficult to know what the instrumental
noise
level is to better than perhaps a factor of two by conventional methods.  Tinto
et al. \cite{tinto} have emphasized the value of using the symmetrized Sagnac calibration
 to determine the total  gravitational wave signal
for
frequencies of roughly 200 $\mu$Hz to 3 mHz, where the expected level is
above that of the optical path measurement noise.  Our main point is that, after using
Sagnac calibration in properly selected frequency bands where either the
optical path noise or the proof mass noise dominates strongly,
 averaging over a bandwidth comparable with the frequency  
considerably reduces the instrument noise in measurements of
the  smoothed spectral amplitude. This allows better sensitivity  for measurement of
stochastic backgrounds over a larger range of frequencies.

The possible improvement
factor above 5 mHz is up to about $(f \tau/2)^{1/4}$, which equals 20 at
10 mHz, but two types of limitations have to be considered also.  One is due
to the uncertainty in $R_{est}$ at 10 mHz and below.  The other is due to the
similar gravitational wave sensitivities of $\eta$ and $\zeta$ at frequencies of
25 mHz and above.  We estimate that the resulting overall sensitivity
improvement factor for LISA would be between 10 and 20 for frequencies of
about 10 to 25 mHz.

At frequencies below 200 $\mu$Hz, the
improvement factor is about $(f \tau/6)^{1/4}$.  At 100 $\mu$Hz this is a
factor of about 5, so the sum of the galactic and extragalactic backgrounds
could be determined down to   somewhat lower frequencies than otherwise
would be possible.

\section{Sagnac calibration with enhanced  
 high-frequency LISA follow-on mission }

If the LISA mission indeed finds several types of sources involving massive
black holes, there will be strong scientific arguments for follow-on
missions
aimed at achieving considerably higher sensitivity at both lower and higher
frequencies.  Some preliminary discussion of   possible follow-on
missions
has been given by   
Folkner and Phinney\cite{folkner}
and Ungarelli and Vecchio\cite{ungarelli}.  In order to give some
indication of the future background accuracy achievable by calibrating and smoothing,
we consider an illustrative example  of   a high
frequency follow-on mission.

We assume the same basic triangular geometry and 60 degree
ecliptic inclination as for LISA, but the arm lengths  are  
50,000 km
instead of $5\times 10^6$km.  The noise level for the gravitational sensors (i.e. free
mass sensors) is 
  a factor of ten lower than for
 LISA, and the fractional uncertainty in measuring changes in the distances
between the test masses is  
30 times lower than for LISA.  It should be remembered that making the arm
lengths much shorter   also makes the the requirements
on the laser beam pointing stability and on the fraction of a fringe to
which
phase measurements have to be made much tighter.  The shorter antenna
might have the rates of change of the distances between the test masses kept
constant to make the phase measurements on the signals easier, provided that
the required forces on the test masses can be kept free enough of noise.

The extragalactic CWDB background would be
gone above about 0.1 Hz, provided that merger-phase and ringdown radiation
from coalescences are not significant.  At higher frequencies, the binary
background is expected to be almost entirely due to extragalactic neutron
star
binaries and 5 or 10 solar mass black hole
binaries. 
We take
the neutron star binary coalescence rate in our galaxy to be $1\times 10^{-5} {\rm
yr^{-1}}$ , which is a factor of 10 lower than assumed in table 7 of \cite{hils90}.  This
estimate may
still be somewhat on the high side and has a high uncertainty \cite{kalogera}, but we
regard it as giving a plausible estimate of the total gravitational wave
background level, allowing for some additional contribution from black hole
binaries.  We also increase the expected gravitational wave amplitude  by a factor 1.5 to
allow very roughly for eccentricity of the NS-NS binaries\cite{hils91}.  With the ratio of
0.3 between the extragalactic and galactic amplitudes from\cite{kosenko}, this gives an
extragalactic amplitude of $h_{rms,XGNSB}=8\times 10^{-25}  f^{-7/6} {\rm Hz^{-1/2}}$.
The background sensitivity with the
Sagnac calibration gets within a factor of 4 of this extragalactic NS plus BH
binary background at 0.5 Hz. 
  The   follow-on
antenna    would give detailed measurements of the gravitational wave
background spectrum up  to about 100 mHz, as
well
as limits at higher frequencies and  much improved measurements of coalescences of binaries at
cosmological distances containing intermediate mass black holes.

\section{INFORMATION CONCERNING EXTRAGALACTIC ASTROPHYSICAL BACKGROUNDS}

For the LISA mission, the Sagnac calibration approach will make possible
 measurements of the extragalactic CWDB background (XGCWDB)
  at
frequencies from about 5 to 25 mHz.  This is important because it will give
new information on the star formation rate at early times.  Kosenko and
Postnov \cite{kosenko} have investigated the effect of a peak in the star formation
rate
at redshifts of z = 2 or 3 on XGCWDB, with emphasis on the observed
frequency range from 1 to 10 mHz.  However, going to somewhat higher
frequency
would improve the sensitivity to the star formation rate.

The
CWDBs\cite{hils90,benderhils,hilsbender,webbink,webbink2,kosenko,nelemans,schneider,nelemans2}
include He-He, He-CO and CO-CO white dwarf binaries, as well as a few binaries containing the
rarer O/Ne/Mg white dwarfs.  Here He and
CO stand for helium and carbon/oxygen white dwarfs respectively.  Rough
estimates of the comparative rms signal strengths for the first three types
as
a function of frequency are given in Fig. 1 of \cite{benderhils}.  It can be seen that
the frequency cutoffs due to coalescence are different for the different
types, ranging roughly from 15 mHz for the first type to 60 for the third.
This is mainly because the He dwarfs are less massive and larger than the CO
dwarfs.  In addition , the total binary mass ranges for the three types, in
units of the solar mass, are about 0.50-0.75, 0.75-1.45 and 1.45-2.4, which
means that there is a range of coalescence frequencies for each type.

The He-He binaries will contribute the most to determining the star
formation
rate, since their coalescence frequencies at redshifts of 2 or 3 will shift
down into the accurately observable 10 to 25 mHz frequency range and thus
will
change the way in which the XGCWDB varies with frequency.  Information on the
distribution of chirp masses for the different types of CWDBs in our galaxy
can be obtained from the resolved signals above about 3 mHz.  However,
careful
studies will be needed in order to determine the sensitivity of the
resulting
star formation history to factors such as possible differences in the CWDB
chirp mass distribution at earlier times.

The possible high-frequency LISA follow-on mission with Sagnac calibration
would give  
an upper limit to the combined extragalactic NS-NS,
NS-BH and BH-BH binary backgrounds between 0.1 and 1 Hz. For the CO-CO binaries, the highest
frequency signals will come from the merger phase of coalescence and from possible
ringdown of the resulting object, if two conditions are met: that the orbit is nearly
circular before coalescence, and   that a supernova not result.  Even
though
all redshifts will be integrated over, the shape of the upper end of the
CWDB
background seems likely to still give new information on the binary mass
distribution, the coalescence process and the star formation history.

Above 0.1 or 0.2 Hz but below the
range of ground-based detectors, no other astrophysical backgrounds have
been
suggested except those due to extragalactic NS-NS, NS-BH and BH-BH binaries.
Only a crude estimate for the combined background level has been included in
this
paper, and it is highly uncertain.  As has been suggested by a
number
of authors, the   BH-BH binaries may be the dominant source (see
e.g.\cite{schutz,kalogera}).  Higher levels would permit LISA follow-on
observations up to somewhat higher frequencies, where possible confusion
with
a high frequency tail from CO-CO white dwarf merger phase or post-merger
ringdown would be reduced.  Approximate information on the relative
strengths
of the NS and BH binary backgrounds probably will be available from
ground-based observations of the coalescence rates, but probably with only
the
BH-BH coalescences going out to substantial redshifts.  Thus the main new
information from LISA follow-on observations of these backgrounds may be on
the history of the NS binary formation rates.

\section{primordial backgrounds}

We have been characterizing backgrounds by  $h_{rms}^2$, the spectral density of the
gravitational wave strain (also sometimes denoted $S_h$).  For cosmology, we are interested 
in sensitivity in terms of the broadband energy density of an isotropic, unpolarized,
stationary background, whose cosmological importance is characterized by
\begin{equation}
\Omega_{GW}(f)\equiv  
\rho_c^{-1} {d\rho_{GW}\over d \ln f}
={4\pi^2\over 3 H_0^2}f^3 h_{rms}^2(f)
\end{equation}
where we   adopt  units of the critical density $\rho_c$.
The  broadband energy density per $e$-folding of frequency, $\Omega_{GW}(f)$, is
thus related to the   rms strain spectral density 
 by\cite{maggiore}
\begin{equation}
{h_0^2\Omega_{GW}\over 10^{-8}}
\approx
\left({h_{rms}(f )\over 2.82\times  10^{-18}{\rm \ Hz^{-1/2}}}\right)^2
\left({f\over 1{\rm mHz}}\right)^3,
\end{equation}
where $h_0$ conventionally denotes Hubble's constant in units of 
$100{\rm km\ s^{-1}\ Mpc^{-1}}$.
In these units, the main sources of instrumental and astrophysical noise are
summarized  schematically  in Figure 
\ref{fig: noises}.

Primordial backgrounds  can be produced by a variety of classical mechanisms
producing relativistic macroscopic or mesoscopic energy flows 
  at $T\ge 100$GeV, whose only  
 observable relic is  a gravitational
wave background\cite{maggiore,kosowsky,apreda}.   
Because the gravitational radiation processes are not perfectly efficient, the  total energy
density
$\int d\ln f\Omega_{GW}$ in gravitational waves must be  less than that in the thermal
relativistic relic particles (photons and three massless neutrinos)
where the ``waste heat'' resides today,
$\Omega_{rel}h_{0}^{2}=4.17\times 10^{-5}  T_{2.728}^4$,
where $h_{0}$ refers to the Hubble constant. (The integrated
density  is already limited
by nucleosynthesis arguments to less than about 0.1 of this value because
of the effect on the expansion rate.)  
It is interesting to pursue stochastic backgrounds as far as possible below this
maximal level  
since most predicted effects, for example waves from  even strongly 
first-order phase transitions, are at least several orders of magnitude weaker.

The  spectrum of the background  conveys  
information  on    early stages of cosmic history. 
Classical processes    typically produce backgrounds   covering a broad band
around a characteristic frequency determined by the scale of the energy flows,
fixed by the gravitational timescale.
  The   band
accessible to the proposed 
 space interferometers, $10^{-5}$ to 1 Hz,
corresponds to the redshifted Hubble frequency from cosmic temperatures between about
100 GeV  and $10^4$ TeV--- often thought to include processes such as baryogenesis and
supersymmetry breaking, and possibly also
 activity in new extra dimensions\cite{hoganbrane1,hoganbrane2}.  
We adopt the point of view  that 
 it is interesting to explore new regions of 
frequency and amplitude for broad-band backgrounds, regardless of
 theoretical justifications for a particular scale.
 We present in Figure \ref{fig: regions} a summary of the
likely accessible  parameters (frequency and amplitude) for  primordial backgrounds,
optimistically taking account of the improvements suggested here,
both for LISA and the  illustrative   high-frequency successor 
considered earlier.

 A much more ambitious goal often cited is detection of
  gravitational waves expected from the
quantum fluctuations of the graviton field during inflation. These occur at all
frequencies up to the  redshifted Hubble frequency from the inflationary epoch (which
may  exceed $10^{12}$ Hz),  but are 
in general much weaker than the classical sources; a na{\"\i}ve estimate is that 
$\Omega_{GW,inflation}\approx 
h_{inflation}^2\Omega_{rel}$
where $h_{inflation}\approx (H_{inflation}/M_{Planck})$ is the amplitude of  tensor
metric quantum fluctuations on the Hubble scale, and
 $H_{inflation}$ is the Hubble constant during inflation. From the 
microwave background anisotropy we estimate that on large
scales,  
$h_{inflation}
\approx (\delta T/T)_{tensor}\le 10^{-5}$. Unless the spectrum is 
``tilted'' in an  unexpected direction (larger $H_{inflation}$ on 
smaller scales, which inflate last), this is an upper limit on the 
quantum effects and is a rough estimate of where gravitational
wave data set limits on ``generic'' models of inflation. The corresponding
$h_0^2\Omega_{GW}\approx 10^{-15}$ is  about ten orders of magnitude below the maximal
classical level, and well below the astrophysical binary noise.

The problem of separating primordial backgrounds from binary backgrounds depends to
some extent on how different the spectra are. From general scaling
arguments\cite{hoganbrane1},  
   classical phase transitions, where the radiation is emitted
over a  short period of time, tend to
generate  spectra with a steep low frequency  limit, scaling
like 
 $\Omega_{GW}\propto f^{7}$ to $f^{6}$. The 
high frequency limit in some models (involving defects such as light
cosmic strings or Goldstone waves,  or brane displacement
modes) may be scale-invariant, $\Omega_{GW}\propto$ constant; in 
phase transitions it falls off 
at least as fast as 
 $\Omega_{GW}\propto f^{-1}$ and can be even steeper.  
Even though these  processes have characteristic frequencies, the 
primordial spectrum is quite broad and is not expected to have sharp features that would
stand out as diagnostics.   At  frequencies above 100 mHz, where the astrophysical
confusion  
background is mainly from  neutron star and black hole  binaries (for which
 the main energy
loss
  is gravitational radiation),  it obeys the scaling
$\Omega_{GW} \propto f^{2/3}$.  At lower frequencies 
the dominant XGCWDB spectrum departs from this
due to redshift and various nongravitational effects on the binary population, as
discussed earlier;   in the 
10 to 100 mHz range the dominant XGCWDBs are  predicted to closely mimic   a scale-free
spectrum.
Depending on the situation,  spectral features may or may not clearly distinguish  a
primordial component.

For the LISA mission with Sagnac
        calibration, the estimated astrophysical background apparently
 can be detected with
        S/N$\approx$  10 from 10 to 25 mHz, but because of the uncertainties in modeling
XGCWDB, it is not clear whether a primordial contribution  
         could be detected for a level much less than $h_0^2\Omega_{GW}\approx
10^{-10}$.
        On the other hand,   the range of frequencies over which
        the primordial background search can be carried out 
at this level is substantial, especially 
if we include the possible LISA follow-on mission  discussed
earlier (see Figure 4).      
With the high-frequency antenna, the results   with Sagnac calibration
may reach  a   level below 
$h_0^2\Omega_{GW}\approx 10^{-11}$
at a frequency   above 0.1 Hz where there is a drop in the
astrophysical backgrounds.

Recently Cornish and Larson\cite{cornishlarson} and Cornish\cite{cornish}
 have discussed
further the use of cross-correlation of signals from two similar antennas to search
for a primordial background.  In particular, Cornish and Larson suggest that
such antennas with roughly 1 AU baselines and operating near
$ 1\times 10^{-6}$ Hz might
be able to reach a sensitivity for $ h_o^2 \Omega_{GW}$ of about $10^{-14}$.  With
Sagnac calibration, a low-frequency LISA follow-on mission could in principle
reach a similar primordial background sensitivity with a single antenna, as
well as provide additional valuable information on MBH-MBH binaries.

\section{appendix: $\delta E $ for high-frequency estimator}
  The uncertainty $\delta E $ in estimating $\langle E_k\rangle$ can be estimated from the
relation
\begin{equation}
        \delta E^2 = \langle E_k^2\rangle - \langle E_k\rangle^2.
\end{equation}
We will be dealing with $ E_k$, but drop the subscript $k$ for now.  
Expanding Eq. (\ref{eqn: estimator}) into noise and wave parts, 
\begin{eqnarray}
3E &=& \{[|X_n|^2+|Y_n|^2+|Z_n|^2]-[8 \sin^2(2 \pi fL)]|\zeta_n|^2\}_1      \\ \nonumber 
& & +\{[(X_n)(X_{GW})^* + (Y_n)(Y_{GW})^* + (Z_n)(Z_{GW})^*] - [8 \sin^2(2
\pi fL) (\zeta_n)(\zeta_{GW})^*] + c.c.\}_2 \\        \nonumber       &
&+\{[|X_{GW}|^2+|Y_{GW}|^2+|Z_{GW}|^2]-[8 \sin^2(2 \pi fL)]|\zeta_{GW}|^2\}_3\hfil
\end{eqnarray}

For frequencies where $S_{GW,ave}$ is small compared with $S_{n,ave}$ and where
$|\zeta_{GW}|^2$ can be neglected,
\begin{equation}
        E \approx  (1/3)\{ \}_1 + (1/3)[|X_{GW}|^2 + |Y_{GW}|^2 + |Z_{GW}|^2];
\end{equation}
since the noise terms average to zero,
\begin{equation}
        \langle  E \rangle \approx S_{GW},
\end{equation}
(as promised for the estimator, by design), and
since their spectral density dominates the GW terms,
\begin{equation}
        \langle  E^2 \rangle \approx  (1/9)\langle [\{ \}_1]^2\rangle.
\end{equation}

   We assume that the lengths of the three arms for LISA are nearly equal
to their average value $L$.  From the definitions of $X, Y, Z$ and $\zeta$ in
ref. \cite{estabrook}:
\begin{equation}
        X = (y_{32} - y_{23})[e^{3i \omega L} - e^{i \omega L}]
                + (y_{31} - y_{21})[e^{2i \omega L} - 1],
\end{equation}
\begin{equation}
        Y = (y_{13} - y_{31})[e^{3i \omega L} - e^{i \omega L}]
                + (y_{12} - y_{32})[e^{2i \omega L} - 1],
\end{equation}
\begin{equation}
        Z = (y_{21} - y_{12})[e^{3i \omega L} -e^{i \omega L}]
                + (y_{23} - y_{13})[e^{2i \omega L} - 1],
\end{equation}
\begin{equation}
        \zeta = [(y_{32} + y_{21} + y_{13}) - (y_{23} + y_{31} + y_{12})]e^{i \omega L}.
\end{equation}
Then:
\begin{equation}
        |X|^2 =  4 \sin^2(\omega L) [|y_{32}-y_{23}|^2 + |y_{31}-y_{21}|^2
                       + \{(y_{32}-y_{23}) (y_{31}^* - y_{21}^*)e^{i \omega L} + c.c. \}],
\end{equation}
etc.  From such expressions, it can be shown that:
\begin{equation}
        \langle  E^2 \rangle \approx  (32/9)\sin^4(\omega L)   \{\langle  \Xi_1\rangle
           + [5 + 4 \cos (\omega L)]\langle  \Xi_2\rangle + [6 - 2 \cos (2 \omega L)]\langle 
\Xi_3\rangle\},
\end{equation}
where
\begin{equation}
   \Xi_1 =
|y_{32}|^2|y_{23}|^2+|y_{21}|^2|y_{12}|^2+|y_{13}|^2|y_{31}|^2       
+|y_{31}|^2|y_{21}|^2+|y_{12}|^2|y_{32}|^2+|y_{23}|^2|y_{13}|^2,
\end{equation}
\begin{equation}
   \Xi_2 =
|y_{23}|^2|y_{31}|^2+|y_{31}|^2|y_{12}|^2+|y_{12}|^2|y_{23}|^2
+|y_{32}|^2|y_{13}|^2+|y_{13}|^2|y_{21}|^2+|y_{21}|^2|y_{32}|^2,
\end{equation}
\begin{equation}
   \Xi_3 =
|y_{32}|^2|y_{31}|^2+|y_{13}|^2|y_{12}|^2+|y_{21}|^2|y_{23}|^2.
\end{equation}

   It is clear from these expressions that the uncertainty $\delta E$ in $E$
depends on the instrumental noise levels in the six main signals, rather
than just on their average.  Assuming, however, that they are all equal
and uncorrelated gives the following estimates:
\begin{equation}
   \langle  E^2 \rangle ~=  (64/3)\sin^4(\omega L) [9 + 4\cos(\omega L) - \cos(2\omega L)]
                       \langle S_y^{optical path}\rangle^2,
\end{equation}
\begin{equation}
        \delta E ~< 16 \sin^2(\omega L)\langle S_y^{optical path}\rangle.
\end{equation}
In general, even if the instrumental
noise levels are unequal, it can be shown that $\delta E$ is less than
$ \sqrt{9/8} \eta^2$, so that $\lambda \stackrel{<}{\sim} \sqrt{9/8}\approx 1$
for frequencies from 5 to 25 mHz.
\acknowledgements

It is a pleasure to thank John Armstrong, Frank Estabrook and Massimo Tinto
for extensive discussions of their laser phase noise correction method and
particularly for emphasizing the possibility of using the Sagnac observable
to correct for instrumental noise.  We also thank the following for valuable
discussions of what could be achieved by possible LISA follow-on missions:
Sterl Phinney, Bill Folkner, Ron Hellings, Bernard Schutz, Carlo Ungarelli,
Alberto Vecchio, Karsten Danzmann, and Neil Cornish. This work was supported
at the University of Washington by NSF and at the University of Colorado by NASA. 

 {}

\begin{figure}[htbp] 
\centerline{\epsfig{height=6in, file=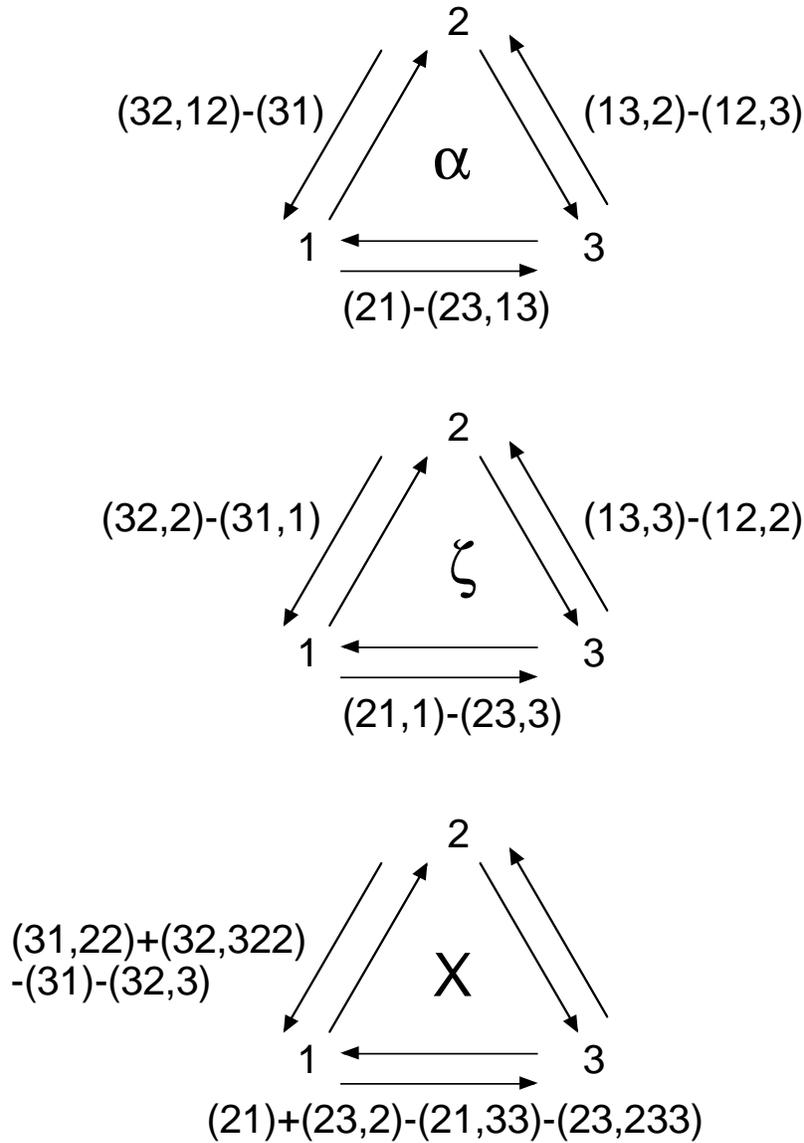}}
\vspace{0.25in}
\caption{ \label{fig: sagnacpicture}
Illustration of the signal combinations discussed in the text. The numbers
labelling each pair of  arrows
correspond to the subscript labels of signals in the notation of Armstrong et al.:
``12,3'' for example refers to $y_{12,3}$, the signal  traveling on the side opposite 
spacecraft 1, received by spacecraft 2 (from spacecraft 3),  with a 
time delay corresponding to the light travel time along the side opposite 3.
The $\beta$ and $\gamma$ observables correspond to cyclic permutations of the
indices for $\alpha$. The symmetrized Sagnac
observable $\zeta$ is very similar to the round-trip-difference observables
$\alpha,\beta,\gamma,$ except that for $\zeta$ all the signals are compared with almost
the same time delays, leading to a minimal sensitivity to low-frequency gravitational waves.
The $X$ observable is based on   a Michelson interferometer using only two
sides, but is the difference in signals at two times separated by
approximately the the round trip 
travel time on one arm. The  $Y$ and $Z$ observables
are equivalent to $X$ but based on  the other spacecraft pairings. }
\end{figure}
 
\begin{figure}[htbp] 
\centerline{\epsfig{height=6in, file=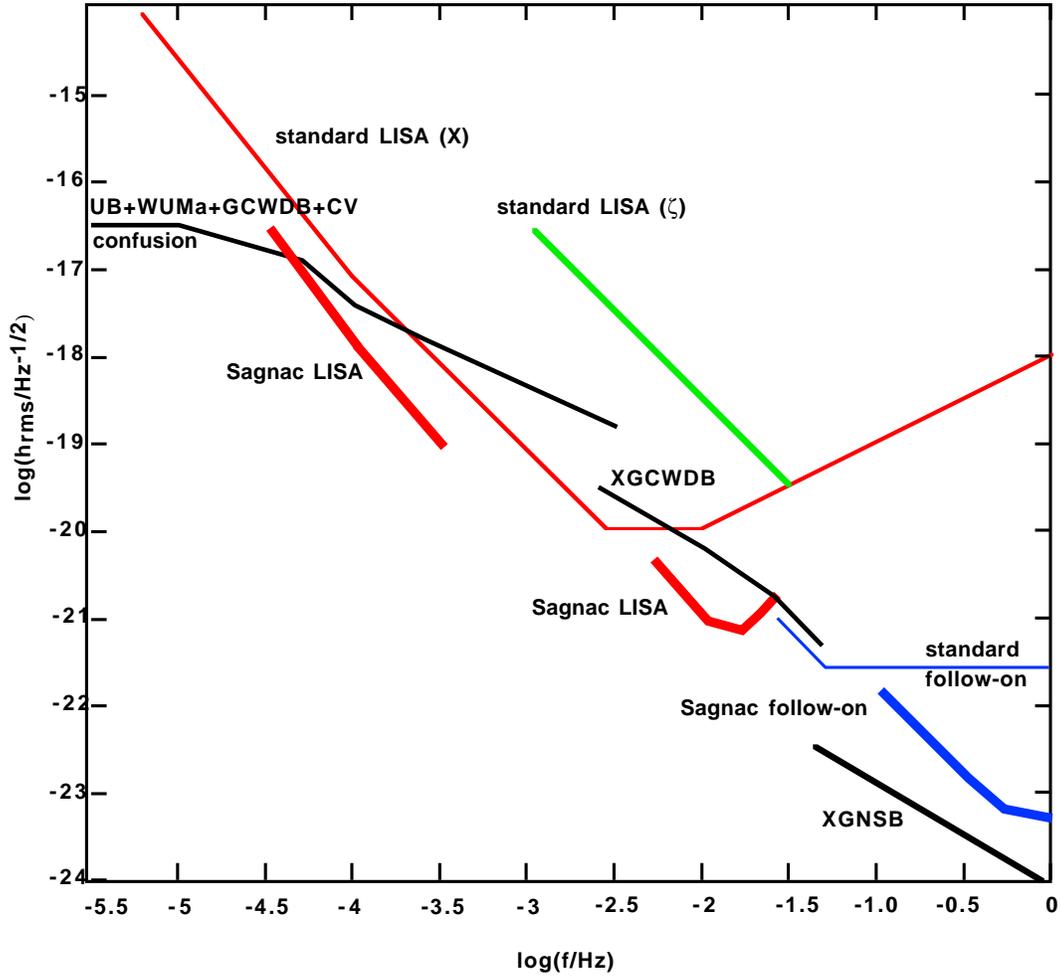}}
\vspace{0.25in}
\caption{ \label{fig: strain}   Instrument sensitivity
in terms of rms strain per $\sqrt{\rm Hz}$, to broad band backgrounds,
assuming a one year integration. The ``standard''
S/N=1 levels in one frequency resolution element, 
  for LISA  and for the
shorter-baseline follow-on mission described in the text, are
shown  as lighter lines. 
The sensitivity is shown for both the (standard) Michelson observable $X$ and 
the symmetrized Sagnac observable $\zeta$. The levels
theoretically attainable with  Sagnac calibration and averaging over bandwidth $f/2$
are shown in bold lines. The Sagnac estimator loses its advantage
at high frequencies where $\zeta$ is no longer insensitive to gravitational waves;
the analytic form for the  estimator discussed here is also inefficient
 at  frequencies where the proof mass noise and optical path noise are comparable.
At low frequencies where proof mass noise dominates, another  analytic 
form yields a significant improvement in sensitivity, which allows the 
confusion background to be measured to lower frequencies. 
  Estimated
astrophysical backgrounds are shown  for Galactic binaries, extragalactic white dwarf
binaries, and extragalactic neutron star or black hole binaries.
}
\end{figure}

\begin{figure}[htbp] 
\centerline{\epsfig{height=6in, file=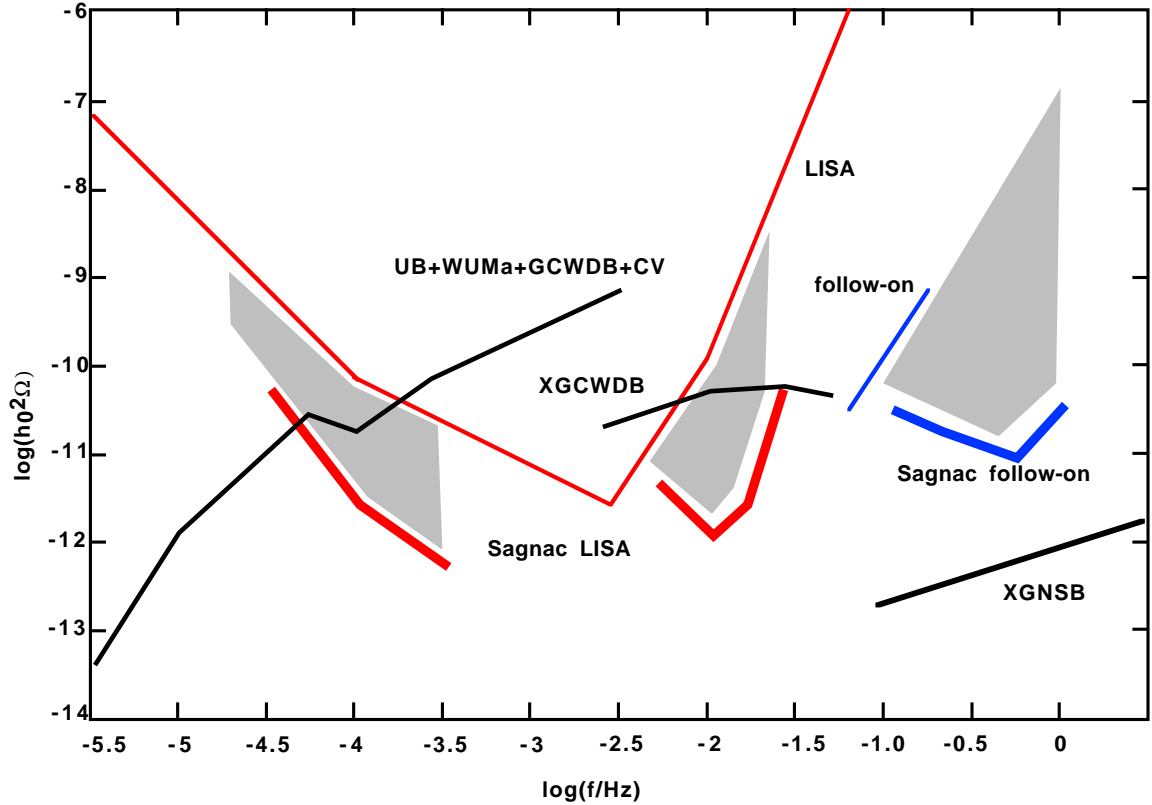}}
\vspace{0.25in}
\caption{ \label{fig: noises} Noise levels are shown in terms of  
the equivalent energy density of  an isotropic
stochastic  background.  Units are the energy
density per factor $e$ of frequency, in units of the critical density, normalized for
Hubble constant $h_0=1$. Where applicable, a one-year integration is assumed. 
The sum UB+WUMa+GCWDB represents
 the estimated confused background from
the sum of  unevolved Galactic binaries,
W Ursa Majoris binaries and white dwarf binaries. These estimates are
uncertain by about a factor of 10 in $\Omega$.
 The confusion noise   level
drops abuptly above the frequency where almost all
 Galactic binaries can be fitted
out. Extragalactic white-dwarf  binaries ``XGCWDB'' create a stochastic confusion
noise which cannot be eliminated.    At still higher frequencies 
above about 0.1 Hz,  the white dwarfs
coalesce, leaving only the confusion background from  extragalactic
 neutron star   binaries and stellar-mass black hole binaries (XGNSB). The
LISA instrument noise limit (S/N=1) after one year is shown, both the traditional
narrow-band sensitivity and the broad-band sensitivity allowed by  Sagnac calibration
and discussed here. The shaded regions  show  the main  areas for 
improvement   possible  from using Sagnac calibration.
The Sagnac technique allows a significantly improved
measurement of a  low resolution spectrum
of  the confusion background   with LISA  both at low frequencies $\approx 0.1$mHz
and at higher frequencies to
$\ge$ 20 mHz, including an accurate measurement with LISA of the
extragalactic white dwarf binary confusion background.
The Sagnac sensitivity limit  for the   smaller-baseline 
  follow-on mission  is
shown for the parameters discussed in the text; in this case the Sagnac technique
offers a more substantial overall improvement in sensitivity.}
\end{figure}

\begin{figure}[htbp] 
\centerline{\epsfig{height=6in, file=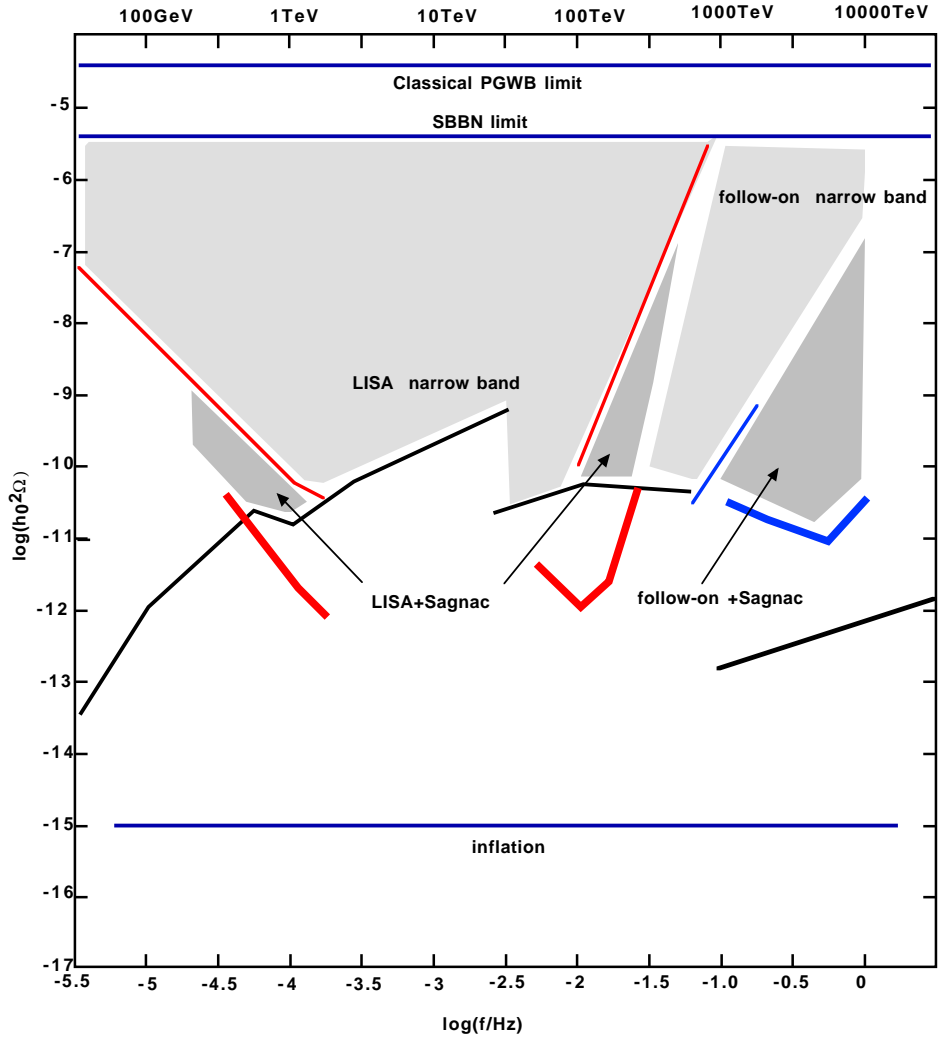}}
\vspace{0.25in}
\caption{ \label{fig: regions} Regions of new parameter space
for primordial backgrounds opened up by   proposed experimental
setups and data analysis strategies.  
Scale   on the top axis shows  the 
cosmic temperature for which  classical
waves generated at the Hubble frequency and redshifted to the present
yield the observed frequency on the bottom axis.
Several characteristic energy densities are shown:
Classical primordial gravitational wave background limit
(PGWB) shows the sum of energies of
photons and massless neutrinos, the maximal level expected for primordial backgrounds;
``SBBN'' denotes the maximum level consistent with Standard Big Bang
Nucleosynthesis (both of these for a background with $\Delta f= f$); and ``inflation''
denotes a typical, untilted, scale-free inflation-generated spectrum, at the maximum
level consistent with the background radiation anisotropy. Shaded regions
lie above both instrument noise and binary confusion backgrounds, where
primordial backgrounds can be detected. The darker-shaded regions show  the
extra benefit (for primordial background measurements) of Sagnac calibration with both
missions.  }
\end{figure}

\end{document}